\documentclass[aps,psfig,showpacs,groupedaddress]{revtex4-1}
\usepackage{color}
\usepackage{graphics}
\usepackage{epsfig}

\newcommand{\bee}{\begin{equation}}
\newcommand{\ene}{\end{equation}}
\newcommand{\beea}{\begin{eqnarray}}
\newcommand{\enea}{\end{eqnarray}}

\begin{document}
\title{Exact propagating nonlinear singular disturbances in strongly coupled dusty plasmas}
\author{Amita Das}
\affiliation{Institute for Plasma Research, Bhat , Gandhinagar - 382428, India }
\author{Sanat Kumar Tiwari}
\affiliation{Institute for Plasma Research, Bhat , Gandhinagar - 382428, India }
\author{Predhiman Kaw}
\affiliation{Institute for Plasma Research, Bhat , Gandhinagar - 382428, India }
\author{Abhijit Sen}
\affiliation{Institute for Plasma Research, Bhat , Gandhinagar - 382428, India }
\date{\today}
\begin{abstract} 
The nonlinear longitudinal response of a strongly coupled dusty plasma system is
analytically investigated using the Generalized Hydrodynamic (GHD) model.
It is shown that the Galilean invariant form of the model does not have
soliton solutions, but can support a variety of nonlinear singular (yet conservative) solutions 
like shock structures of zero strength and propagating 
solutions with cusp like singularities.  These novel entities should be detectable 
in numerical simulations and experiments studying large amplitude longitudinal excitations in 
such systems.  

 

\end{abstract}
\maketitle 
When micron sized dust is sprinkled in a plasma, the dust particles  acquire a negative charge and shielding clouds large enough to balance thermal electron currents to ion thermal currents.
The shielded dust particles  surrounded by their ion and electron Debye clouds thus form an additional charged component of the plasma. 
If the dust density is large enough, the Yukawa interaction energy between two shielded dust particles may readily exceed their kinetic energies
putting the dust component into a strongly coupled state whereas the electrons and ions still remain in a weakly 
coupled regime \cite{Morfill2009,khrapak_morfill_2009}. Strongly coupled dusty plasmas may be found in many physical 
situations ranging from planetary astrophysical plasmas 
\cite{goertz_89} to plasmas in industrial environment \cite{selwyn_1991} and even plasmas at the boundaries of thermonuclear 
fusion devices \cite{Krasheninnikov_04}. 
Such plasmas have also been produced 
 in the laboratory for basic experiments  where one has observed interesting phenomena such as formation and melting of crystalline structures
\cite{chu_94,thomas_1994,thomas_morfill} and excitation of collective modes in the crystalline as well as liquid phases of the dust component 
\cite{pieper_96,melandso_1996,Nunomura_2000,pramanik_2002_prl,pintu_pla}.
There have also been a number of theoretical investigations of strongly
coupled dusty plasmas using a variety of techniques \cite{rosenberg_1997,wang_97,kawghd98,murillo_98}. Among them the Generalized Hydrodynamic (GHD) model
approach \cite{kawghd98}, that augments the usual hydrodynamic description by a phenomenological visco-elastic term to 
account for the strong coupling correlations, has been fairly successful in describing the linear collective properties
of a strongly coupled dusty plasma system. For example, the existence of transverse shear waves in the liquid state 
and strong coupling induced modifications in the dispersion properties of dust acoustic waves were predicted by
the GHD model \cite{kawghd98} and subsequently verified in laboratory experiments \cite{pramanik_2002_prl,pintu_pla} and 
molecular dynamics simulation \cite{ohta_hamaguchi_prl}. 
The application of  the 
GHD model in the nonlinear high amplitude regime has however remained quite sparse and limited, 
primarily due to technical difficulties in analyzing the equations.  Unlike  the weak coupling regime, the equations
cannot be reduced to a KdV like paradigmatic equation that has soliton solutions. In fact, as we will demonstrate,
the nonlinear GHD model cannot support soliton solutions. Our present investigations show, that under some realistic approximations and 
in the weak amplitude limit, the nonlinear GHD equations
can be reduced to the Hunter-Saxton equation, an integrable nonlinear PDE with a very different and distinct class of exact nonlinear solutions. 
In contrast to the usual KdV solitons, these solutions have a spatial singularity and yet retain conservative 
properties including many soliton like collision characteristics.
 In the more general case, for arbitrary amplitude excitations, we show that the GHD equations have nonlinear solutions 
in the form of  propagating cusp like structures
for the velocity perturbations. Our results therefore offer a new paradigm for analysing nonlinear excitations in strongly coupled dusty plasmas 
and point to new directions of research for numerical simulation and experimental explorations.\\
 

The basic equations of the Generalized Hydrodynamic (GHD) model describing the dusty plasma medium in one dimension are \cite{kawghd98},
\begin{equation}
\frac{\partial n }{\partial t} + \frac{\partial \left(n u \right)}{\partial x} = 0
\label{nldce}
\end{equation}
\bee
\left[ 1+\tau_m  \frac{d}{d t} \right] 
\left[ \frac{d u} {d t}
 + \left(\frac{T_d}{T_i}\right) \frac{1}{z_d n } 
\frac{\partial P}{\partial x} - \frac{\partial \phi}{\partial x} \right] =
\eta^{*}\frac{\partial ^{2} u }{\partial x^{2}}
\label{ndae2}
\ene
\begin{equation}
\frac{\partial^{2}\phi}{\partial x^{2}} = \left[n + \mu_{e}exp(\sigma_{i}
\phi) - \mu_{i}exp(-\phi)\right]
\label{nldpe}
\end{equation}
Here,  $n$ is the dust density, $u$  is   the  dust velocity along $x$ and $\eta^{\ast}$ is the viscosity coefficient.  
The evolution of $u$ 
 incorporates  characteristic traits associated with strongly coupled matter through a relaxation 
 parameter $\tau_m$. 
 For time scales longer than $\tau_m$ the medium behaves like a viscous liquid,  whereas  at shorter time scales the 
memory  effect  persists and the system shows solid like elastic properties.  
The operator  $d/dt = \partial/\partial t + u \partial/\partial x$ denotes  the total time derivative. 
It may be mentioned here that the convective 
term appearing with coefficient $\tau_m$ in Eq.(\ref{ndae2}) has often been dropped as a simplification in 
several earlier studies  \cite{shukla_noconvect,mamun_pre_2009}. However, this term ensures 
the Gallilean invariance of the equation and ought to be retained.   
The scalar potential  $\phi $   is determined from the Poisson's equation (Eq.(\ref{nldpe})). 
The  time scales associated with the evolution of the dust being long, 
 the electron and ion species  are assumed to satisfy the   Boltzman distribution 
 $ n_e = \mu_{e}exp\left(\sigma_{i}\phi\right) $ and $n_i = \mu_{i}exp\left(-\phi\right)  $ 
 and have been accordingly specified in the Poisson's equation.The dust pressure is modelled by a simple equation of state $P=k_B n T_d$. Above equations have been written in terms of normalized fields 
 described in one of our earlier works \cite{Veer_sanat2010}.
In this paper we concentrate on the strongly coupled limit 
$k \sqrt{\tau_m \eta^{\ast}} >> 1$, where $k$ is the inverse scale length of the solution. We shall primarily use analytic methods 
for drawing conclusions. 

If  $\eta^{\ast}/\tau_m >> C_{da}^2$ ($C_{da}$ being 
the dust acoustic speed), i.e. when the elastic wave dominates the dust acoustic speed, 
one can ignore the contribution from the scalar potential $\phi$ and the thermal contribution due to $P$ in the momentum equation. 
Physically, this is the regime when elastic coefficients due to correlations dominate 
 over  Boltzmann screening and thermal dispersion effects. The dusty plasma medium is, however, still in a fluid molten state with 
 no lattice formation. In this limit the dust fluid is governed by the following simplified equation: 
\begin{equation}
\left[\frac{\partial}{\partial t} + u \frac{\partial}{\partial x}\right]\left[ \frac{\partial u}{\partial t} + 
u \frac{\partial u}{\partial x} \right] =
\frac{\eta^{*}}{\tau_m}  \frac{\partial ^{2} u}{\partial x^{2}}
\label{ndae4}
\end{equation}
It should be emphasized that compressional velocity perturbations in the dust fluid will still  produce density disturbances, 
which in turn will be shielded by electrons 
and ions producing potential perturbations. The inequality at the beginning of this paragraph ensures that the reaction back of these 
driven disturbances on the momentum equation is negligible. 
Physically eq.(\ref{ndae4}) contains dispersion free linear elastic waves that are supported by the correlation driven 
elasticity coefficient and nonlinear contributions through inertial effects appear through the convective terms.
 In principle, linear wave dispersion may be introduced  
through a $k$ dependent form of $\tau_m$ \cite{Murillo2000_prl}; 
here we assume that this effect is small.  Note that the second convective derivative which arose through constraints of Galilean 
invariance is playing a crucial role in the nonlinear dynamics. 
This equation can also model plastic flow deformation disturbances in solids undergoing failure through severe stresses.  

In the weakly nonlinear regime 
Eq.(\ref{ndae4}) can be subjected to 
a reductive perturbation analysis, by expanding,
$u  = \lambda + \epsilon u^{(1)} + \epsilon^{2} u^{(2)} +  \ldots$
%
and
$v = \epsilon^{3/2} (v^{(1)} + \epsilon v^{(2)}  + \ldots)$
where $v = (\partial/\partial t + u \partial/\partial x)u$.
\noindent
Further, using the stretched variables, $\xi = \epsilon^{1/2} \left(x-\lambda t\right)$, 
$\tau = \epsilon^{3/2} t$, taking $\lambda = \sqrt{\eta^{\ast}/\tau_m}$, and retaining terms  upto second order for the $u$ and $v$ fields, we can 
obtain the following single equation in the variable $u^{(1)}$ (rewritten below without the superscript),
\bee
(u_{\tau} + u u_{\xi} )_{\xi} = \frac{1}{2} u_{\xi}^2
\label{ndae7}
\ene
The left hand side equated to zero is the nonlinear equation for dispersionless waves with the convective nonlinearity giving indefinite
 steepening of 
waves which can lead to wave breaking or form shocks, or solitons depending on whether nonlinearity, viscous dissipation (Burgers equation) 
or dispersion 
(Korteweg de Vries equation) dominates the physics of steepened waves. Here the extra convective derivative 
nonlinearity of the simplified generalized 
hydrodynamic model equation (\ref{ndae4}) is responsible for the nonlinear term on the right side of equation (\ref{ndae7}). 
This term dramaticaly changes the character of the equation and the nature of its solutions.

Equation (\ref{ndae7}) is the so-called Hunter-Saxton equation, which has been derived earlier \cite{hunter_saxton} for 
director fields in liquid crystals, where the positional disorder of 
polymer molecules gives the medium fluid properties whereas the orientational order due to correlations gives them crystal like properties. 
It is also the 
high frequency limit of the Cammasa- Holm equation \cite{cammasa_holm} , which has been derived to describe the nonlinear dynamics 
and wave breaking of shallow water waves.
These equations belong to a new class of equations which can be derived from variational principles in more than one non equivalent forms. 
They typically have an infinite number of conservation laws and possess singular solutions with infinite derivatives.
 If these solutions are propagating, 
they pass through each other undisturbed, except for a phase shift, somewhat like solitons.

We now recapitulate some properties \cite{hunter_zheng} of the Hunter Saxton equation and its solution, 
which are of relevance to our problem, Firstly, integrating Eqn. (\ref{ndae7}) over $\xi$ 
we note that because of the positive definite value of the integral on the right side, if the solutions leave one boundary unperturbed, 
the other boundary is perturbed, 
thus showing the impossibility of smooth periodic or isolated solutions with undisturbed boundaries. 
Secondly, a Lagrange variable treatment for the derivative $u_{\xi}$
shows that the velocity derivative blows up in a finite time. This indicates that the nonlinear disturbances 
will lead to a wave breaking like behaviour in a finite time. 
However, since correlations lead to elasticity, we find wave breaking phenomenon with a difference.  
This is best illustrated by the exact solution \cite{hunter_saxton} below.

Hunter Saxton (HS) equation has step like piecewise continuous and non - smooth solutions. A single step solution may be written as 
\begin{eqnarray}
u(\xi,\tau; \alpha, \beta ) &=& U(\xi,\tau;\alpha)  \hspace{0.3in} if  \hspace{0.3in} \tau \le 0  \nonumber \\
   &=& U(\xi,\tau;\beta)  \hspace{0.3in} if \hspace{0.3in} \tau \ge 0 
\label{ndae8}
\end{eqnarray}
where $\alpha$ and $\beta$ are positive constants with condition $\beta \le \alpha$ and,
\begin{eqnarray}
U(\xi,\tau; \alpha ) &=& -\alpha \tau   \hspace{0.3in}  -\infty < \xi \le -\frac{\alpha \tau^2}{2}  \nonumber \\
   &=&  2\frac{\xi}{\tau}  \hspace{0.3in}  -\frac{\alpha \tau^2}{2}   < \xi < 0  \nonumber \\
  &=& 0 \hspace{0.3in} 0 \le \xi < \infty
\label{ndae9}
\end{eqnarray}
Also $U(\xi,0,;\alpha) = 0$ so that $U$ is a continuous function of $\xi$ and $\tau$. Mathematically, the solution is a weak solution of the HS equation satisfying 
the condition 
\begin{equation}
\int \left[  \Phi_{\xi \tau} u + \frac{1}{2} \Phi_{\xi}  u^2 - \frac{1}{2}   \Phi u_{\xi}^2\right]  d\xi d\tau = 0 
\label{ndae10}
\end{equation}
for arbitrary test function $\Phi$; this is weakly admissible if 
\begin{equation}
(u_{\xi}^{2})_{\tau} +(uu_{\xi}^2)_{\xi} = 2 (\beta - \alpha) \delta(\tau) \delta(\xi) 
\label{ndae11}
\end{equation}
From Eqn.(\ref{ndae11}) we note that conservation of $\int u_{\xi}^2 d\xi$ is strictly valid if $\beta = \alpha$. 
For any other value $\beta < \alpha$, the solution is 
dissipative but still weakly admissible. The solutions are non propagating in the wave frame travelling to the right with the 
linear phase velocity $\sqrt{ (\eta^{\ast}/\tau_m)}$.  

Unlike the inviscid Burger equation, where the step becomes vertical (shock solution) and acquires a finite steady value consistent with conservation laws, the HS equation has a step solution 
with a slope and a step size that are time dependent. The remarkable feature of the solution is that as the left and right corners of the linear segment 
having negative slope collide for the creation of a `shock wave' with infinite slope (which may lead to wave breaking) the spatial support in 
real space diminishes to a point and the step size vanishes simultaneously. Since the region of transition diminishes to zero size as the
 step approaches verticality, $\int u_\xi^{2} dx$ can remain conserved. 
In this case no norm is lost and `energy' is conserved. This is unlike normal Burger's like shock wave where the step is constant and 
some energy is converted to heat. In fact at complete verticality, the HS solution has no step and that is why it is sometimes called a
'shock wave of zero strength'. 
\begin{figure}
\includegraphics[height=8.0cm,width=8.0cm]{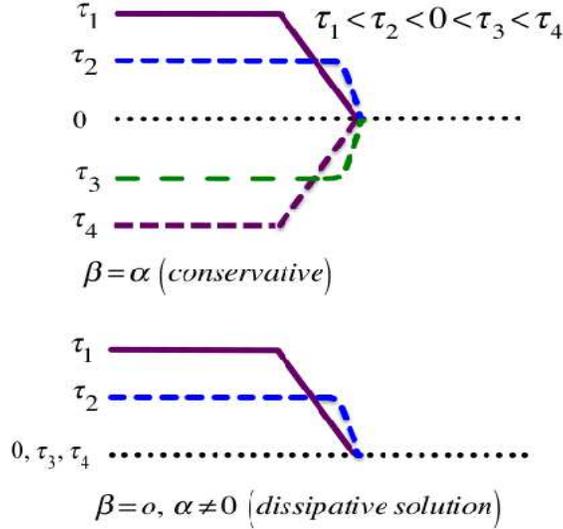}
\caption{Schematic view of time evolution of solution (\ref{ndae8}) of the Hunter Saxton equation
for the conservative and dissipative cases (redrawn from \cite{hunter_saxton}).} 
\label{zero_shock}
\end{figure}
However, we note that such shock waves of zero strength can form either conservative ($\beta = \alpha$) or dissipative 
($\beta < \alpha$ including $\beta = 0$) global solutions. For example, for $\alpha \neq 0$, $\beta = 0$ the disturbance starts with a positive step of $u$ 
on the left at $\tau < 0$ ( see Fig.~\ref{zero_shock}) and then this step goes to zero at $\tau = 0$ when the step becomes vertical. 
Thereafter ($\tau > 0$), the disturbance vanishes from everywhere. This is a dissipative global solution for which the entire energy in the initial 
disturbance damps away and disappears. This is akin to conventional wave breaking, where the infinite slope leads to toppling of the 
wave and conversion of coherent wave energy into chaotic multi stream motions. The conservative global solution, on the other hand, 
corresponds to $\beta = \alpha$ and results in a fresh disturbance with positive slope at $\tau > 0$, where the conserved $\int u_\xi^{2} dx$ energy 
results in a diminishing slope disturbance in a widening region (Fig.~\ref{zero_shock}) as $\tau$ increases . 
This is a remarkable sequel to wave breaking with infinite slope 
at $\tau = 0$, a sequel in which the entire elastic energy of the infinite slope wave trapped  in a region of zero size reappears as a coherent elastic disturbance. This is only possible due to the presence of 
the RHS of HS equation. The HS equation thus shows evidence for self consistent nonlinear elastic waves supported by correlations. which steepen indefinitely but peter out 
in strength before they reach verticality; these are weakly nonlinear waves which steepen and want to break but cannot do so because of strong coupling and correlations. Physically, one may picture the global dissipative and conservative solutions described above as longitudinal disturbances along a spring attached to a wall. Imagine a compressional disturbance coming towards the wall, steepening and becoming infinitely compressed at the wall. The subsequent behaviour can be either inelastic (global dissipative solution) with the entire energy in the disturbance dissipated at the wall (say, because of plastic failure of the spring) and nothing returning back or elastic with a longitudinal disturbance of equal magnitude returning from the wall.

We now examine the arbitrary amplitude Eqn.(\ref{ndae4}), for some special exact solutions. 
We wish to explore the possibility of singular solutions with singular derivatives. For this purpose we introduce a functional of 
$x - \beta t$ as the independent variable, viz. 
\begin{equation}
f = \kappa(x -\beta t)+ G(f)
\label{ndae12}
\end{equation}
and assume that $u$ is a function of $x,t$ only through the combination $f$. This means solution $u(x,t)$ is time independent in a frame moving with velocity 
$\beta$. If $G(f)$ is chosen as a constant, it simply acts as a phase term. If $G$, on the other hand is chosen as a function of $f$, then $f$ becomes a functional of $x -\beta t  $; this dependence  can 
be made implicit if $G$ depends on $f$ through $u$. This is indeed what we shall do, so as to derive an exact cusp like solution of $u(x,t)$ with 
infinite derivatives at the maximum. From Eqn.(\ref{ndae12}) we get,
\begin{equation}
(u - \beta) \frac{d}{df} \frac{u - \beta}{1-G_f} \frac{du}{df} = \frac{\eta^{\ast}}{\tau_m} \frac{d}{df} \frac{1}{1-G_f} \frac{du}{df}
\label{ndae14}
\end{equation}
Introducing $u-\beta =\sqrt{\eta^{\ast}/\tau_m} U$ and making the choice $1-G_f = \sqrt{\eta^{\ast}/\tau_m} U$, we get the equation
\begin{equation}
U \frac{d^2 U}{df^2} = \frac{d}{df} \frac{1}{U} \frac{dU}{df}
\label{ndae15}
\end{equation}
This equation can be integrated once and rewritten as an energy integral 
\begin{equation}
U_f^2 - \left( \frac{U}{\sqrt{1-U^2}}\right)^2 = 0
\label{ndae16}
\end{equation}
\noindent
where the constant of integration has been absorbed in function $f$ as a multiplication factor and sets the scale of the solution. 
Equation (\ref{ndae16}) is the equation of a zero total energy effective particle moving in a potential energy bowl 
with an inverse parabolic form near the origin that blows up 
at $U = \pm 1$. Thus if the effective particle starts from $U =0$ with near zero velocity ($U^{\prime}$),
 it falls towards $U = 1$ slowly at first and then rapidly, till it reaches the ``wall'' (singularity at $U=1$) from where  it is reflected back . It takes an infinite time to climb back to $U = 0$ again. The resulting solution is an isolated soliton like solution with a cusp at the maximum $U=1$. The solution showing this property is 
\begin{eqnarray}
U &=& sech{(f+\sqrt{1-U^2})}
\end{eqnarray}

This solution is illustrated in Fig {\ref{cuspon} and shows infinite derivatives and a cusp singularity at 
 $U=1$ (that is, $u= \beta + \sqrt {(\eta^{\ast}/\tau_m}$). The phase speed and the scale size of theese solutions are not 
directly determined by the maximum amplitude. The nature of singularity can be explored by expanding the solution around $U = 1$
 and demonstrating that $U^{\prime} \approx f^{-1/3}$.  These cuspon like solutions are steady propagating solutions which are 
singular at a point and are dithering at the wave breaking amplitude.  
Physically, such solutions might arise when smooth nonlinear waves acquire amplitudes close to wave breaking, 
but because of conservation laws squeeze the infinite derivative region to a point with a finite elastic energy content.  
That this is indeed so, can be ascertained  from the integral  $\int U^{\prime 2} df = \int U^{\prime} dU =
 \int U dU/\sqrt{1-  U^2}  = 1$ in normalized units. Such finite energy content singular solutions may have special 
stability properties. 
It may be worthwhile to look for an infinite number of conservation laws for the exact Eqn.(\ref{ndae4}), for if they are found then
two such solutions could propagate through each other without distortions, like solitons.
In fact, the stability and accessibility of these solutions from arbitrary initial 
conditions is a topic worthy of further investigation.

\begin{figure}
\includegraphics[height=8.0cm,width=8.0cm]{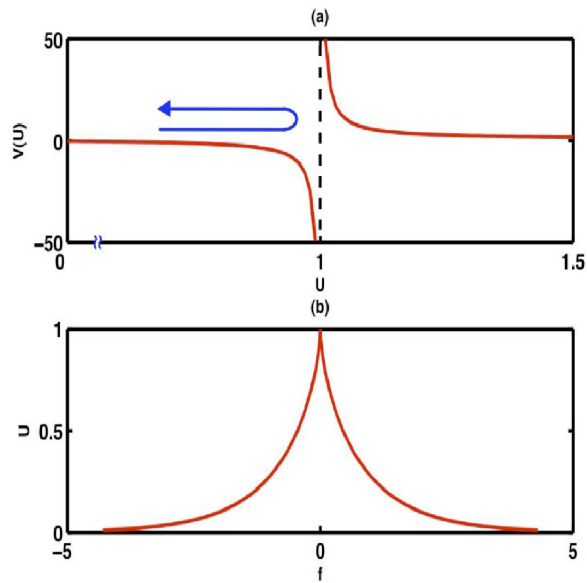}
\caption{Form of potential energy ``bowl'' and zero energy ``particle'' orbit corresponding to a cuspon solution.} 
\label{cuspon}
\end{figure}

In conclusion, we have investigated the nonlinear longitudinal disturbances in a strongly coupled dusty plasma 
using the description of generalized hydrodynamic equations. 
 We work in the limit of very strong coupling where the correlation induced elasticity dominates the dispersion relation 
for linear longitudinal waves. For the weakly nonlinear case we have found that the nonlinear waves are described by the 
Hunter Saxton equation. This equation has exact non smooth solutions which are in the nature of time dependent 
steps and slopes propagating at the speed of linear elastic waves. We have considered a single step solution and shown 
how it gives rise to a shock wave of zero strength, which either produces a global dissipative solution or a global 
conservative solution. The former leads to a damping of the coherent wave energy like in normal wave breaking events 
whereas the latter leads to the reappearance of a step like elastic disturbance which carries off the energy trapped in 
the zero size singularity region associated with the shock wave of zero strength. For the arbitrary amplitude case we have 
discovered the existence of cuspon like solutions, viz. propagating longitudinal isolated disturbances with cusp like singularities at the maximum. 
It would be very worth while to look for cuspons, shock waves of zero strength and global dissipative and conservative solutions in 
laboratory and simulation investigations of nonlinear disturbances in strongly coupled dusty plasmas. It is possible that the cuspon 
solutions are related to cusp like solutions already observed in some dusty plasma experiments \cite{teng_2009}.

{\bf{Acknowledgement: } }
This work was  financially supported by  DAE-SRC grant with sanction number:  2005/21/7-BRNS/2454. 

%

\end{document}